# Casimir-Lifshitz forces and plasmons in a structure of two dielectric rods: Green's function method of electrodynamics

M. V. Davidovich

Saratov National Research State University named after N. G. Chernyshevsky,

410012 Saratov, Russia

davidovichmv@info.sgu.ru

A new model for calculating the Casimir-Lifshitz force per unit length for two dielectric rods is proposed, based on the Green's function method of classical electrodynamics and the Lorentz model for permittivity. For metal rods, it is proposed to use the Drude-Smith model. It is shown that the origin of the force is associated with the fluctuation excitation of slow coupled long-wave surface polaritons. A simple model of the dispersion forces for two CNTs is proposed.

Fluctuating electromagnetic fields are responsible for such important physical phenomena as thermal radiation, radiative heat transfer, van der Waals interactions at small distances between molecules and nanoclusters, including van der Waals friction, the Casimir effect, and the Casimir-Lifshitz forces between bodies [1–16]. The interaction between nanoobjects began to be studied with the work of Casimir [2]. It can have the character of attraction [2,3,15,16] or repulsion [4,5,15,16]. The Lifschitz theory for arbitrary temperature was constructed for objects extended in depth, separated by a flat vacuum gap [3]. The extension means the total attenuation of the incident radiation deep into the structure, and therefore Lifshitz considered only the force acting from the side of the gap on the surface of the half-space. Dielectric rods and carbon nanotubes (CNTs) are not extended objects. In a number of works, the Lifshitz theory was developed using a quantum-field approach based on the temperature Green functions (GF's) of quantum statistical physics [7,8], as well as applied to more complex configurations [9–16].

In this paper, we consider an approach to the determination of the Casimir-Lifshitz linear forces for dielectric rods and CNTs based on the tensor GF's of electrodynamics and the Maxwell tension tensor (MTT), the form of which is known for vacuum. The approach differs from the approaches using tensor GF of structures [1,15], since well-known free space GF's are used. The dispersion forces for this configuration have not been considered in the literature.

Let us consider two parallel infinite rectangular dielectric rods with dimensions $2a_n$, $2b_n$, $n=1,2$, located at a distance $d$ along the $z$ axis, Fig. 1. The polarization currents $\mathbf{J}(\omega,\mathbf{r}) = \mathbf{J}^p(\omega,\mathbf{r}) = i\omega\varepsilon_0(\varepsilon(\omega,\mathbf{r})-1)\mathbf{E}(\omega,\mathbf{r})$ in the rods are expressed in terms of an electric field and excite a secondary electromagnetic field. The self-consistent system of equations is the system of volume integral equations. Waves are possible in the considered structure. Free waves (without excitation by external sources) are dissipative plasmon-polaritons (PP) along the *z*-axis. We consider forced PP caused by fluctuations, when fluctuation currents are considered to be incident sources. Such forced (non-dissipative) PP is characterized by a real propagation constant $\gamma$, along which the spectral decomposition (integration) takes place.

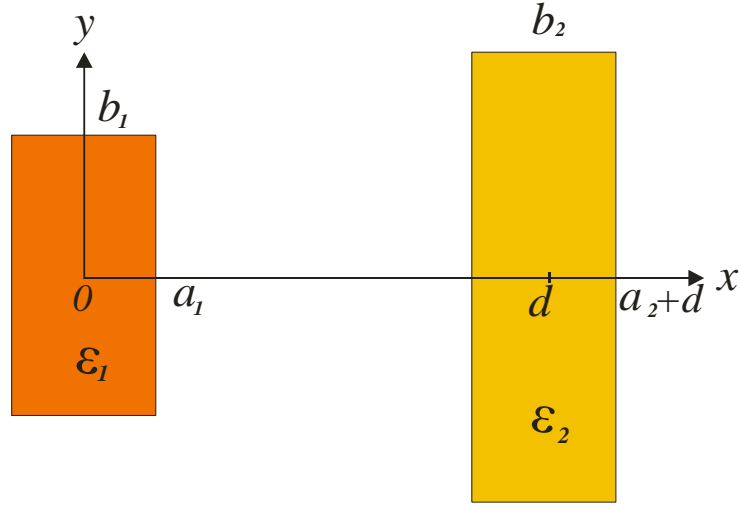

Fig. 1. Configuration of two interacting dielectric rods

We use the following simplifications of the model. First – we consider the rods to be infinite along the *z*-axis and parallel. Second – we consider only the longitudinal components of the polarization current $\mathbf{J}_n(\omega,\mathbf{r}) = \mathbf{z}_0 J_{nz}$, $J_{nz} = i\omega\varepsilon_0(\varepsilon_n(\omega)-1)E_{nz}(\omega,\mathbf{r})$. For small transverse dimensions, the longitudinal polarization significantly exceeds the transverse one, which justifies this approximation. In this case, we consider the rods to be thin, i.e. the fields in them are assumed to be independent of the transverse coordinates. This is true when the considered wavelengths in the rods are substantially larger than the transverse dimensions. With transverse dimensions of less than a few tens of nm, almost the entire absorption spectrum of the substance satisfies this condition. This allows us to move from a three-dimensional model to a quasi-one-dimensional model, significantly simplifying the calculations, but taking into account the real configuration. The quantity $\sigma_{nz} = i\omega\varepsilon_0(\varepsilon_n(\omega)-1) = \sqrt{\varepsilon_0/\mu_0}\,k_0\xi_{nz}$ where dimensionless quantities $\xi_{nz} = i(\varepsilon_n(\omega)-1)$ are introduced can be considered as the conductivity. At the same time $\sigma_{nx} = \sigma_{ny} = 0$, therefore, we omit the longitudinal index: $\sigma_{nz} = \sigma_n$. Third – each rod is described by the longitudinal permittivity $\varepsilon_{nzz} = \varepsilon_n$ associated with the conductivity $\sigma_n$, assuming that the density of the longitudinal polarization current $J_{nz} = J_n$ ($n=1,2$) is constant over the cross-section of each CNT, and the transverse currents are not taken into account. Although there is a transverse polarization, its contribution to the excited fields is small compared to the longitudinal one. Fourth, we consider the distance $d$ to be significantly larger than the transverse dimensions. Actually, this allows us to neglect the current densities of polarization $J_{nx}, J_{ny}$, which must be taken into account in a strict theory. Accounting $J_{ny}$ is necessary if we consider the interaction of infinite plates along the $z$ and $y$ axes. At the same time, for narrow such plates, the current $J_{1x}$ can be neglected. This allows us to consider the interaction of the plates approximately. However, this is a more complex problem: taking into account the four components $J_{nx}, J_{ny}$, $n=1,2$ leads to a system of four equations with four unknowns, which is much more difficult when considered analytically than for the two equations in the case of rods. Fluctuating currents $J_{1z}^0(\mathbf{r}) = J_1^0(\mathbf{r})$, $J_{2z}^0(\mathbf{r}) = J_2^0(\mathbf{r})$ occur in the dielectric. If there are no free charges, these are also the polarization currents (note that the free charge currents are also included in the polarization currents). For them, we also neglect the transverse fluctuations. We write Maxwell's spectral equations in the form

$$\nabla \times \mathbf{H} = ik_0 \varepsilon Z_0^{-1} \mathbf{E} + \mathbf{z}_0 \left[ J_{1z}^0(\mathbf{r}) + J_{2z}^0(\mathbf{r}) \right], \quad \nabla \times \mathbf{E} = -ik_0 Z_0 \mathbf{H}. \tag{1}$$

The force acting on a body with a surface $S$ can be introduced in several ways. We will define it as in [3] in terms of the MTT, which is equal to the minus sign of the spatial part of the energy-momentum tensor( EMT), which in the SI system has the form

$$T_{\alpha\beta} = \varepsilon_0 E_\alpha E_\beta + \mu_0 H_\alpha H_\beta - \delta_{\alpha\beta} \left[ \varepsilon_0 \mathbf{E}^2 + \mu_0 \mathbf{H}^2 \right]/2. \tag{2}$$

The Greek indices run through the values of $x, y, z$, and the same indices are summed. EMT determines the pressure, and the tension force on the body is opposite and equal to

$$\mathbf{f} = -\int_S \hat{T}(\mathbf{r}_s) \mathbf{n}(\mathbf{r}_s) d^2 r. \tag{3}$$

For random fields, take the autocorrelation function $\langle \hat{T}(\mathbf{r}_s) \rangle$. For real values, it means stochastic averaging. For complex amplitudes, the correlation function (hereinafter referred to as correlation) $\langle E_\alpha, E_\beta \rangle$ means the statistical averaging of a value $E_\alpha E_\beta^*$. Only the $x$-axis force acts on the first rod

$$F_x = -\int_{-b_1}^{b_1} \int_{-l}^{l} \left[ \langle T_{xx}(a_1, y, z) \rangle - \langle T_{xx}(-a_1, y, z) \rangle \right] dy dz. \tag{4}$$

Here we have introduced a force acting on the length $L=2l$, assuming quite reasonably (this will be shown later) that the tension is the same over the entire length:

$$F_x = -L \int_{-b_1}^{b_1} \left[ \langle T_{xx}(a_1, y, z) \rangle - \langle T_{xx}(-a_1, y, z) \rangle \right] dy. \tag{5}$$

The contour integral gives the linear force $f_x = F_x / L$. In our case

$$\langle T_{xx} \rangle = \varepsilon_0 \frac{\langle E_x, E_x \rangle - \langle E_y, E_y \rangle - \langle E_z, E_z \rangle}{2} + \mu_0 \frac{\langle H_x, H_x \rangle - \langle H_y, H_y \rangle}{2}, \quad (6)$$

$$\langle T_{xy} \rangle = \varepsilon_0 \langle E_x, E_y \rangle + \mu_0 \langle H_x, H_y \rangle. \quad (7)$$

The tensor has the dimension of the energy density J/m$^3$ or N/m$^2$, so it determines the pressure (negative tension). We will use spectral values rather than time values. In the harmonic Maxwell equations (1), the dimension of the vectors **E** and **H** is the same as in the non-stationary ones. However, the spectral densities from time fields have a different dimension, the denominator of which additionally includes the frequency. Accordingly, spectrum integration gives time fields and time correlations. The force will be determined by frequency integration, using spectral values. The values of the type $\varepsilon_0 \langle E_x(\mathbf{r}, \omega), E_y(\mathbf{r}, \omega) \rangle$ have the dimension of action. In our thin rod approximation for the electric field in the zero approximation, we can leave one correlation $\langle E_z, E_z \rangle$, which simplifies the result, but we will use all three.

As in [3], we will look for the solution of equations (1) as the sum of the partial solution of inhomogeneous equations (1), which we denote $\mathbf{E}^0$, $\mathbf{H}^0$ ($J_n^0 \neq 0$), and the general solution of homogeneous equations $\mathbf{E}^d$, $\mathbf{H}^d$ (at $J_n^0 = 0$). Here, the index $d$ essentially corresponds to the fact that these are the fields of diffraction (multiple scattering) of radiation from fluctuating sources: $\mathbf{E} = \mathbf{E}^0 + \mathbf{E}^d$, $\mathbf{H} = \mathbf{H}^0 + \mathbf{H}^d$. Consider the linear force acting on the first rod. We only take into account the component $T_{xx}$ on the two sides $x = \pm a_0$, since the other components do not give a contribution. The spectral electric and magnetic fields in the structure under consideration are defined in terms of the Green tensor functions of free space as [17]

$$\mathbf{E}(\mathbf{r}) = \mathbf{E}^0(\mathbf{r}) + \int_{V_1+V_2} \hat{\Gamma}^e(\mathbf{r}, \mathbf{r}') \mathbf{z}_0 J_z(\mathbf{r}') d^3 r', \quad (8)$$

$$\mathbf{H}(\mathbf{r}) = \mathbf{H}^0(\mathbf{r}) + \int_{V_1+V_2} \hat{\Gamma}^h(\mathbf{r}, \mathbf{r}') \mathbf{z}_0 J_z(\mathbf{r}') d^3 r'. \quad (9)$$

Here, the integration goes over the volumes of the rods, and the fluctuation fields are caused by current fluctuations:

$$\mathbf{E}^0(\mathbf{r}) = \int_{V_1+V_2} \hat{\Gamma}^e(\mathbf{r}, \mathbf{r}') \mathbf{z}_0 J_z^0(\mathbf{r}') d^3 r', \quad (10)$$

$$\mathbf{H}^0(\mathbf{r}) = \int_{V_1+V_2} \hat{\Gamma}^h(\mathbf{r}, \mathbf{r}') \mathbf{z}_0 J_z^h(\mathbf{r}') d^3 r'. \quad (11)$$

These fields in turn excite the diffraction fields and create full fields. The complete field (8) satisfies the impedance conditions of the form $E_{nz} = \sigma_{nz}^{-1} J_{nz}$ (in fact, it is a continuity condition of $E_{nz}$). The notation (8) and (9) formally follows from the representation $ik_0 \varepsilon Z_0^{-1}\mathbf{E} = ik_0 Z_0^{-1}\mathbf{E} + \mathbf{z}_0 J_{nz}$ of the first term in the right-hand side of the first equation (1), which corresponds to the Maxwell equations in a vacuum with two types of sources: $J_{nz}^0$ and $J_{nz}$. Next, we will omit the $z$ indices, and assume the $n$ index to be 1 or 2, depending on whether $z$ belongs to the first or second rods. Tensor GF are defined in terms of the scalar FG:

$$\hat{\Gamma}^e(\mathbf{r},\mathbf{r}') = \frac{k_0^2 \hat{I} + \nabla \otimes \nabla}{i\omega\varepsilon_0} G(\mathbf{r}-\mathbf{r}'),$$

$$\hat{\Gamma}^h(\mathbf{r},\mathbf{r}') = \hat{R}(\mathbf{r}) G(\mathbf{r}-\mathbf{r}'),$$

where $(\nabla \otimes \nabla)_{\alpha\beta} = \partial_\alpha \partial_\beta$, and the matrix $\hat{R}$ corresponds to the "rotor" operator $\nabla \times$ and has the form

$$\hat{R}(\mathbf{r}) = -\hat{\varepsilon}_{\alpha\beta\gamma} \partial_\gamma = \begin{bmatrix} 0 & -\partial_z & \partial_y \\ \partial_z & 0 & -\partial_x \\ -\partial_y & \partial_x & 0 \end{bmatrix}.$$

Here a completely antisymmetric Levi-Civitta tensor $\hat{\varepsilon}_{\alpha\beta\gamma}$ is introduced. In our case, only the components $\Gamma_{xz}^e = (i\omega\varepsilon_0)^{-1}\partial_x\partial_z$, $\Gamma_{yz}^e = (i\omega\varepsilon_0)^{-1}\partial_y\partial_z$, $\Gamma_{zz}^e = (i\omega\varepsilon_0)^{-1}(k_0^2 - \partial_z^2)$, $\Gamma_{xz}^h = \partial_y$, $\Gamma_{yz}^h = -\partial_x$ are used. It is convenient to take the spectral representation of GF in the form [17]

$$G(\mathbf{r}) = \int_{-\infty}^{\infty}\int_{-\infty}^{\infty} \frac{\exp(-i\gamma z - i\beta y - \kappa|x|)}{8\pi^2 \kappa} d\beta d\gamma. \qquad (12)$$

Here $\kappa = \sqrt{q^2 - k_0^2}$, $q^2 = \gamma^2 + \beta^2$, and in the case of $\gamma^2 + \beta^2 < k_0^2$ we take a branch of the root $\kappa = i\sqrt{k_0^2 - q^2}$. We will also use the following equivalent spectral representation of GF:

$$G(\mathbf{r}) = \int_{-\infty}^{\infty}\int_{-\infty}^{\infty}\int_{-\infty}^{\infty} \frac{\exp(-i\alpha x - i\beta y - i\gamma z)}{8\pi^3} d\alpha d\beta d\gamma. \qquad (13)$$

Spectral representations of the tensor GF are obtained by substitutions, for example, $\partial_z \to -i\gamma$ etc. On the right side of the left (first) rod, we have $E_x(a_1,y,z) = E_x^0(a_1,y,z) + E_x^d(a_1,y,z)$, where

$$E_x^d(a_1,y,z) = \int_{V_2}\int_{-\infty}^{\infty} \frac{\gamma\exp(-i\gamma(z-z') - i\beta(y-y'))}{-8\omega\varepsilon_0 \pi^2} d_0(x-x') J_2(x',y',z') d\gamma d\beta dx' dy' dz'. \qquad (14)$$

Here the function $d_0(x-x') = \exp(-\kappa|x-x'|)$ is introduced, and the fluctuation component of the field is

$$E_x^0(a_0,y,z) = \int_{V_2}\int_{-\infty}^{\infty} \frac{\gamma\exp(-i\gamma(z-z') - i\beta(y-y'))}{-8\omega\varepsilon_0 \pi^2} d_0(x-x') J_2^0(x',y',z') d\gamma d\beta dx' dy' dz'.$$

Moving on to integrals over positive domains, we have

$$E_x^0(a_0, y, z) = 4i \int\int\int\limits_{V_2} \int\limits_0^\infty \frac{\gamma \sin(\gamma(z-z'))\cos(\beta(y-y'))}{8\omega\varepsilon_0 \pi^2} d_0(x-x') J_2^0(x', y', z') d\gamma d\beta dx' dy' dz'.$$

In (14), we excluded the term corresponding to the volume integral of the first rod, since it does not depend on the distance $d$ and determines the self-action. However, the magnitude $J_2$ on the second body due to diffraction depends on $J_1^0$ and $J_1$. When differentiating the module, a function $\text{sgn}(a_0 - x') = -1$ occurred. Always $d_0(a_0 - x') = \exp(\kappa(x' - a_1))$, therefore, the differentiation gives $+1$. Next, we use the Fourier transform:

$$J_n^0(\alpha, \beta, \gamma) = \int\limits_{-a_0}^{a_0} \int\limits_{-a_0}^{a_0} \int\limits_{-\infty}^{\infty} J_n^0(x, y, z) \exp(i\alpha x + i\beta y + i\gamma z) dx dy dz, \quad (15)$$

$$J_n^0(x, y, z) = \frac{1}{(2\pi)^3} \int\limits_{-\infty}^{\infty} J_n^0(\alpha, \beta, \gamma) \exp(-i\alpha x - i\beta y - i\gamma z) d\alpha d\beta d\gamma, \quad (16)$$

and similar transformations for the fields. We denoted the triple integral in (16) by single symbol. We will do this further, defining multiplicities by differentials. We will also denote $J_{nz}^0(\alpha, \beta, \gamma) = J_{nz}^0(\mathbf{k})$, $\exp(i\alpha x + i\beta y + i\gamma z) = \exp(i\mathbf{kr})$. The average values of the current densities are zero: $\langle J_{nz}^0(\mathbf{r}) \rangle = 0$, and their correlations are given by the relation [18]

$$\langle J_n^0(\mathbf{r}), J_m^0(\mathbf{r}') \rangle = -i\delta_{nm}(2\pi)^{-1}\omega\varepsilon_0 \Theta(\omega, T)(\varepsilon_{nzz}^* - \varepsilon_{nz})\delta(\mathbf{r} - \mathbf{r}'), \quad (17)$$

or $\langle J_n^0(\mathbf{r}), J_m^0(\mathbf{r}) \rangle = \delta_{nm} F_n(\omega, T) = \delta_{nm} \omega\varepsilon_0 \Theta(\omega, T) \varepsilon_n'' / \pi$, $F_n(\omega, T) = \omega\varepsilon_0 \Theta(\omega, T) \varepsilon_n'' / \pi$. Here $\varepsilon_n'' = -\text{Im}(\varepsilon_n)$ and the average energy of a quantum oscillator at a temperature $T$ is denoted as [18]:

$$\Theta(\omega, T) = \frac{\hbar\omega}{2} + \frac{\hbar\omega}{\exp(\hbar\omega / k_B T) - 1} = \frac{\hbar\omega}{2} \coth\left(\frac{\hbar\omega}{2k_B T}\right).$$

To perform (17), it is necessary and sufficient for the spectral correlations to be performed $\langle J_n^0(\mathbf{k}), J_m^0(\mathbf{k}') \rangle = (2\pi)^3 \delta_{nm} \delta(\mathbf{k} - \mathbf{k}') \omega\varepsilon_0 \Theta(\omega, T) \varepsilon_n'' / \pi$. However, this cannot be strictly performed and is performed the better the larger the area occupied by the dielectric (see below). In the general case $\langle J_n^0(\mathbf{k}), J_m^0(\mathbf{k}') \rangle = 2\pi\delta_{nm}\delta(\gamma - \gamma')\tilde{F}_n(\alpha, \alpha, \beta, \beta, \mathbf{a}, \mathbf{b})\omega\varepsilon_0 \Theta(\omega, T) \varepsilon_n'' / \pi$, where a function $\tilde{F}_n$ of dimensions $\mathbf{a} = (a_1, a_2)$, $\mathbf{b} = (b_1, b_2)$, and spectral parameters is introduced.

Let consider the correlation $\langle E_x^0(a_1, y, z), E_x^0(a_1, y, z) \rangle$, given the delta-correlation of $\langle J_{2z}^0(\mathbf{r}), J_{2z}^0(\mathbf{r}') \rangle$ in the form (17):

$$\langle E_x^0(a_1, y, z), E_x^0(a_1, y, z) \rangle = F_2(\omega, T) \cdot$$

$$\cdot \int\int\limits_{V_2} \int\limits_{-\infty}^{\infty} \gamma\gamma' d_0 d_0'^* \frac{\exp(i(\gamma' - \gamma)z + i(\beta' - \beta)y)\exp(-i(\gamma' - \gamma)z' - i(\beta' - \beta)y' - (\kappa + \kappa'^*)(x' + a_1))}{(8\omega\varepsilon_0 \pi^2)^2} d\gamma d\beta d\gamma' d\beta' d^3 r'.$$

Here we have presented the components in the form of decompositions (15). Integrating over $z'$, we also get $2\pi\delta(\gamma-\gamma')$ and the result of integrating over $\gamma'$:

$$\langle E_x^0(a_1,y,z), E_x^0(a_1,y,z)\rangle = 2\pi F_2(\omega,T)\cdot$$

$$\cdot \int_{d-a_2}^{d+a_2}\int_{-b_2}^{b_2}\int_{-\infty}^{\infty}\gamma^2 \frac{\exp(i(\beta'-\beta)y)\exp(-i(\beta'-\beta)y'-(\kappa+\kappa'^*)(x'-a_1))}{(8\omega\varepsilon_0\pi^2)^2}d\gamma d\beta d\beta' dx'dy'.$$

Integrating over $y'$, we get

$$\langle E_x^0(a_1,y,z), E_x^0(a_1,y,z)\rangle = 4\pi F_2(\omega,T)\cdot$$

$$\cdot \int_{d-a_2}^{d+a_2}\int_{-\infty}^{\infty}\gamma^2 \frac{\sin((\beta'-\beta)b_2)\exp(i(\beta'-\beta)y-(\kappa+\kappa'^*)(x'-a_1))}{(8\omega\varepsilon_0\pi^2)^2(\beta'-\beta)}d\gamma d\beta d\beta' dx'.$$

Here $\kappa'^* = (\gamma^2+\beta'^2-k_0^2)^{1/2*}$, so in the area $\beta'^2 > k_0^2 - \gamma^2$ we have $\kappa'^* = \kappa'$. Also in the area $\beta^2 > k_0^2 - \gamma^2$ the value $\kappa$ will be real. In the remaining regions, these values are imaginary. The area $\beta^2 < k_0^2 - \gamma^2$ and $\beta'^2 < k_0^2 - \gamma^2$ should be excluded, since the value $d_0 d_0^*$ does not depend on $d$. Therefore, the correlation is valid and even positive. Indeed, taking the complex conjugation of the integral and making a substitution $\beta \leftrightarrow \beta'$, we see that the result does not change. Formally, you can put the function Re before the integral. Integrating over $x'$ in the area of the second body and denoting

$$d_0^2 = 2\sinh((\kappa+\kappa'^*)a_2)\frac{\exp(-(\kappa+\kappa'^*)(d-a_1))}{(\kappa+\kappa'^*)},$$

we have

$$\langle E_x^0(a_1,y,z), E_x^0(a_1,y,z)\rangle = 4\pi F_2(\omega,T)\int_{-\infty}^{\infty}d_0^2\gamma^2 \frac{\exp(i(\beta'-\beta)y)\sin((\beta'-\beta)b_2)}{(8\omega\varepsilon_0\pi^2)^2(\beta'-\beta)}d\gamma d\beta d\beta'. \qquad (18)$$

The considered correlation is included in the correlation $\langle E_x(a_1,y,z), E_x(a_1,y,z)\rangle$. It does not depend on $z$. This is a general property of correlations, so when integrating over $z$, infinity occurs, which corresponds to an infinite force. If in (3) we do not perform such integration, but perform integration only along the contour, we get the linear force, which is finite. Integrating by y in the region of the first body, we get

$$\langle E_x^0(a_1), E_x^0(a_1)\rangle = F_2(\omega,T)\int_{-\infty}^{\infty}d_0^2\gamma^2\frac{\sin((\beta'-\beta)b_1)\sin((\beta'-\beta)b_2)}{8\pi(\omega\varepsilon_0)^2(\beta'-\beta)^2}d\gamma d\beta d\beta'. \qquad (19)$$

The integrand in (19) is even in $\gamma$. We integrate over positive values of $\gamma$ and double the result. It is more convenient to exclude from the integration the areas $0 < \gamma^2 < k_0^2 - \beta^2$ and $0 < \gamma^2 < k_0^2 - \beta'^2$ in which there are oscillations. Functions $f^\pm(\beta,\beta')$ also oscillate and have a maximum at $\beta = \beta'$, at the same time $d_0^2 = 0$. By splitting the integral of over the positive and over the negative domains $\beta'$, we obtain even and odd functions of $\beta$ and $\beta'$:

$$f^\pm(\beta,\beta') = \frac{\sin((\beta'-\beta)b_1)\sin((\beta'-\beta)b_2)}{(\beta'-\beta)^2} \pm \frac{\sin((\beta'+\beta)b_1)\sin((\beta'+\beta)b_2)}{(\beta'+\beta)^2}.$$

Both of these functions have the same parity over $\beta$ and $\beta'$, so we take the integration over positive values, doubling the result. Denote $\beta_0^2 = |\gamma^2 - k_0^2|$. We obtain symmetric integrand expressions for correlation

$$\langle E_x^0(a_1), E_x^0(a_1)\rangle = F_2(\omega,T)\int_0^\infty \gamma^2 d\gamma \int_{\beta_0}^\infty \int_{\beta_0}^\infty \frac{d_0^2(\gamma,\beta,\beta')f^+(\beta,\beta')d\beta d\beta'}{2\pi(\omega\varepsilon_0)^2}. \tag{20}$$

The integral with both imaginary $\kappa$ and $\kappa'^*$ should be excluded, since it does not depend on $d$. You can add to the integral (20) a small value of the integral of the oscillating function

$$F_2(\omega,T)\int_0^{k_0} \gamma^2 d\gamma \int_0^{\beta_0} d\beta' \int_{\beta_0}^\infty \frac{2\operatorname{Re}(d_0^2(\gamma,\beta,\beta'))f^+(\beta,\beta')d\beta}{2\pi(\omega\varepsilon_0)^2}.$$

Indeed, calculating the integral by the theorem on the average value at a point $\beta' = \beta$, we also find $d_0^2 = 2a_2$ and the value

$$2a_2 F_2(\omega,T)\int_0^{k_0} \gamma^2 B_0(k_0,\gamma)d\gamma \int_0^{\beta_0} \frac{f^+(\beta,\beta)d\beta}{2\pi(\omega\varepsilon_0)^2}.$$

Here $f^\pm(\beta,\beta) = b_1 b_2 \pm \sin(2\beta b_1)\sin(2\beta b_2)/(2\beta)^2$, $f^+(0,0) = 0$, $f^-(0,0) = 2b_1 b_2$. Since, however, the result is independent of $d$, and it can be omitted. We immediately write out other correlations:

$$\langle E_y^0(a_1), E_y^0(a_1)\rangle = F_2(\omega,T)\int_0^\infty \gamma^2 \int_{\beta_0}^\infty \int_{\beta_0}^\infty \beta\beta' \frac{d_0^2(\gamma,\beta,\beta')f^-(\beta,\beta')d\beta d\beta'}{2\pi\kappa\kappa'(\omega\varepsilon_0)^2}d\gamma,$$

$$\langle E_z^0(a_1), E_z^0(a_1)\rangle = F_2(\omega,T)\int_0^\infty (k_0^2-\gamma^2)^2 \int_{\beta_0}^\infty \int_{\beta_0}^\infty \frac{d_0^2(\gamma,\beta,\beta')f^+(\beta,\beta')d\beta d\beta'}{2\pi\kappa\kappa'(\omega\varepsilon_0)^2}d\gamma,$$

$$\langle H_x^0(a_1), H_x^0(a_1)\rangle = F_2(\omega,T)\int_0^\infty \gamma^2 \int_{\beta_0}^\infty \int_{\beta_0}^\infty \beta\beta' \frac{d_0^2(\gamma,\beta,\beta')f^-(\beta,\beta')d\beta d\beta'}{2\pi\kappa\kappa'}d\gamma,$$

$$\langle H_y^0(a_1), H_y^0(a_1)\rangle = F_2(\omega,T)\int_0^\infty \gamma^2 \int_{\beta_0}^\infty \int_{\beta_0}^\infty \frac{d_0^2(\gamma,\beta,\beta')f^+(\beta,\beta')d\beta d\beta'}{2\pi\kappa\kappa'}d\gamma.$$

Obviously, in these expressions, one can take the integral over $\gamma$ in the limits $(k_0,\infty)$, and then the limits of the integration over $\beta,\beta'$ will be $(0,\infty)$. In this case, we exclude the region $0 < \gamma < k_0$.

From the expression (14), we see that to calculate the correlations of the fields together with $\langle J_n^0, J_n^0\rangle$ the correlations $\langle J_n, J_n\rangle$ and $\langle J_n^0, J_n\rangle$ are necessary. To find them, we use the form FG (13).

Its Fourier spectrum has the form $G(\mathbf{k}) = (\mathbf{k}^2 - k_0^2)^{-1}$. Writing the Fourier component of the vector potential as the sum of the fluctuation and diffraction components $A_z(\mathbf{k}) = A_z^0(\mathbf{k}) + A_z^d(\mathbf{k}) = G(\mathbf{k})(J_{1z}^0(\mathbf{k}) + J_{2z}^0(\mathbf{k}) + J_{1z}(\mathbf{k}) + J_{2z}(\mathbf{k}))$, we find the spectra of the field components

$$E_z = \frac{k_0^2 + \partial_z^2}{i\omega\varepsilon_0} A_z, \ E_x = \frac{\partial_x \partial_z}{i\omega\varepsilon_0} A_z, \ E_y = \frac{\partial_y \partial_z}{i\omega\varepsilon_0} A_z, \ H_x = \partial_y A_z, \ H_y = -\partial_x A_z,$$

or in the spectral representation

$$E_x(\mathbf{k}) = \frac{-\alpha\gamma}{i\omega\varepsilon_0} A_z(\mathbf{k}), \ E_y(\mathbf{k}) = \frac{-\beta\gamma}{i\omega\varepsilon_0} A_z(\mathbf{k}), \ E_z(\mathbf{k}) = \frac{k_0^2 - \gamma^2}{i\omega\varepsilon_0} A_z(\mathbf{k}).$$

The relations for the electric field, expressed in terms of the impedance operator $\hat{\rho}$, can be written as: $\mathbf{E}(\mathbf{k}) = \mathbf{E}^0(\mathbf{k}) + \hat{\Gamma}^e(\mathbf{k})\mathbf{J}(\mathbf{k}) = \hat{\rho}\mathbf{J}(\mathbf{k})$, or $\hat{\rho}\mathbf{J}(\mathbf{k}) = \hat{\Gamma}^e(\mathbf{k})(\mathbf{J}^0(\mathbf{k}) + \mathbf{J}(\mathbf{k}))$, or $\mathbf{J}(\mathbf{k}) = \hat{\sigma}\hat{\Gamma}^e(\mathbf{k})(\mathbf{J}^0(\mathbf{k}) + \mathbf{J}(\mathbf{k}))$. Here the general formulation with tensor DP is used and the inverse specific impedance operator is introduced $\hat{\rho} = \hat{\sigma}^{-1} = (\hat{\varepsilon}(\omega) - \hat{I})^{-1}/(i\omega\varepsilon_0)$. Dividing the current density of the polarization by the sum of two and counting the conductivity of the rods as different, we write

$$\mathbf{J}_1(\mathbf{k}) = \hat{\sigma}_1 \hat{\Gamma}^e(\mathbf{k})(\mathbf{J}_1^0(\mathbf{k}) + \mathbf{J}_2^0(\mathbf{k}) + \mathbf{J}_1(\mathbf{k}) + \mathbf{J}_2(\mathbf{k})),$$

$$(\hat{I} - \hat{\sigma}_1 \hat{\Gamma}^e(\mathbf{k}))\mathbf{J}_1(\mathbf{k}) = \hat{\varsigma}_1 \hat{\Gamma}^e(\mathbf{k})(\mathbf{J}_1^0(\mathbf{k}) + \mathbf{J}_2^0(\mathbf{k}) + \mathbf{J}_2(\mathbf{k})),$$

$$(\hat{I} - \hat{\sigma}_2 \hat{\Gamma}^e(\mathbf{k}))\mathbf{J}_2(\mathbf{k}) = \hat{\sigma}_2 \hat{\Gamma}^e(\mathbf{k})(\mathbf{J}_1^0(\mathbf{k}) + \mathbf{J}_2^0(\mathbf{k}) + \mathbf{J}_1(\mathbf{k})),$$

$$\mathbf{J}_1(\mathbf{k}) = (\hat{I} - \hat{\sigma}_1 \hat{\Gamma}^e(\mathbf{k}))^{-1} \hat{\sigma}_1 \hat{\Gamma}^e(\mathbf{k})(\mathbf{J}_1^0(\mathbf{k}) + \mathbf{J}_2^0(\mathbf{k}) + \mathbf{J}_2(\mathbf{k})),$$

$$\mathbf{J}_2(\mathbf{k}) = (\hat{I} - \hat{\sigma}_2 \hat{\Gamma}^e(\mathbf{k}))^{-1} \hat{\sigma}_2 \hat{\Gamma}^e(\mathbf{k})(\mathbf{J}_1^0(\mathbf{k}) + \mathbf{J}_2^0(\mathbf{k}) + \mathbf{J}_1(\mathbf{k})).$$

Substituting $\mathbf{J}_2(\mathbf{k})$ in the expression for $\mathbf{J}_1(\mathbf{k})$, we find

$$\mathbf{J}_1(\mathbf{k}) = \left(\hat{I} - (\hat{I} - \hat{\sigma}_1 \hat{\Gamma}^e(\mathbf{k}))^{-1} \hat{\sigma}_1 \hat{\Gamma}^e(\mathbf{k})(\hat{I} - \hat{\sigma}_2 \hat{\Gamma}^e(\mathbf{k}))^{-1} \hat{\sigma}_2 \hat{\Gamma}^e(\mathbf{k})\right)^{-1} \cdot$$
$$\cdot (\hat{I} - \hat{\sigma}_1 \hat{\Gamma}^e(\mathbf{k}))^{-1} \hat{\sigma}_1 \hat{\Gamma}^e(\mathbf{k})\left(\hat{I} + (\hat{I} - \hat{\sigma}_2 \hat{\Gamma}^e(\mathbf{k}))^{-1} \hat{\sigma}_2 \hat{\Gamma}^e(\mathbf{k})\right)(\mathbf{J}_1^0(\mathbf{k}) + \mathbf{J}_2^0(\mathbf{k}))$$

(21)

Similarly for second density

$$\mathbf{J}_2(\mathbf{k}) = \left(\hat{I} - (\hat{I} - \hat{\sigma}_2 \hat{\Gamma}^e(\mathbf{k}))^{-1} \hat{\sigma}_2 \hat{\Gamma}^e(\mathbf{k})(\hat{I} - \hat{\sigma}_1 \hat{\Gamma}^e(\mathbf{k}))^{-1} \hat{\sigma}_1 \hat{\Gamma}^e(\mathbf{k})\right)^{-1} \cdot$$
$$\cdot (\hat{I} - \hat{\sigma}_2 \hat{\Gamma}^e(\mathbf{k}))^{-1} \hat{\sigma}_2 \hat{\Gamma}^e(\mathbf{k})\left(\hat{I} + (\hat{I} - \hat{\sigma}_1 \hat{\Gamma}^e(\mathbf{k}))^{-1} \hat{\sigma}_1 \hat{\Gamma}^e(\mathbf{k})\right)(\mathbf{J}_1^0(\mathbf{k}) + \mathbf{J}_2^0(\mathbf{k}))$$

(22)

These relations can be written compactly as $\mathbf{J}_n(\mathbf{k}) = \hat{A}_n (\mathbf{J}_1^0(\mathbf{k}) + \mathbf{J}_2^0(\mathbf{k}))$. To obtain matrices $\hat{A}_n$, in the general case, it is necessary to invert and multiply the three-dimensional matrices included in (19), (20). We have obtained (21) and (22) in general form. In our special case of a single $z$-component (the $z$ index is omitted) $J_1 = \sigma_1 \Gamma^e_{zz}(J_1^0 + J_2^0 + J_1 + J_2)$, $J_2 = \sigma_2 \Gamma^e_{zz}(J_2^0 + J_1^0 + J_2 + J_1)$, so

$$J_1 = \frac{\sigma_1 \Gamma^e_{zz}(J_1^0 + J_2^0 + J_2)}{(1 - \sigma_1 \Gamma_{zz})}, \quad J_2 = \frac{\sigma_2 \Gamma^e_{zz}(J_2^0 + J_1^0 + J_1)}{(1 - \sigma_2 \Gamma_{zz})},$$

$$J_1 = \sigma_1 \Gamma^e_{zz} \frac{J_1^0 + J_2^0}{(1 - \sigma_1 \Gamma^e_{zz})(1 - \sigma_2 \Gamma^e_{zz}) - \sigma_1 \sigma_2 \Gamma^{e\,2}_{zz}} = \sigma_1 \Gamma^e_{zz} \frac{J_1^0 + J_2^0}{1 - (\sigma_1 + \sigma_2)\Gamma^e_{zz}}. \tag{23}$$

In these formulas $\Gamma^e_{zz}(\mathbf{k}) = \sqrt{\mu_0/\varepsilon_0}\,\Gamma(\gamma)/(\mathbf{k}^2 - k_0^2)$, $\Gamma(\gamma) = -i(k_0^2 - \gamma^2)/k_0$. It is issued in the same way for $J_2$. Therefore, we have correlations $\langle J_m(\mathbf{k}), J_n(\mathbf{k'}) \rangle = \langle J_n(\mathbf{k'}), J_m(\mathbf{k}) \rangle^*$ and

$$\langle J_n(\mathbf{k}), J_n(\mathbf{k'}) \rangle = \frac{|\sigma_n|^2 \Gamma^e_{zz}(\mathbf{k}) \Gamma^{e*}_{zz}(\mathbf{k'}) \left[\langle J_1^0(\mathbf{k}), J_1^0(\mathbf{k'}) \rangle + \langle J_2^0(\mathbf{k}), J_2^0(\mathbf{k'}) \rangle\right]}{(1 - (\sigma_1 + \sigma_2)\Gamma^e_{zz}(\mathbf{k}))(1 - (\sigma_1 + \sigma_2)\Gamma^e_{zz}(\mathbf{k}))^*}, \tag{24}$$

$$\langle J_m(\mathbf{k}), J_n(\mathbf{k'}) \rangle = \sigma_m \sigma_n^* \Gamma^e_{zz}(\mathbf{k}) \Gamma^{e*}_{zz}(\mathbf{k'}) \frac{\langle J_1^0(\mathbf{k}), J_1^0(\mathbf{k'}) \rangle + \langle J_2^0(\mathbf{k}), J_2^0(\mathbf{k'}) \rangle}{(1 - (\sigma_1 + \sigma_2)\Gamma^e_{zz}(\mathbf{k}))(1 - (\sigma_1 + \sigma_2)\Gamma^e_{zz}(\mathbf{k}))^*}. \tag{25}$$

The value (25) is proportional $\delta_{mn}$, since the conductivities $\sigma_m$ are different from zero only on the body with the number $m$. Since spectral correlations are considered here, writing them in terms of delta-correlated spatial correlations using (15), we obtain

$$\langle J_2^0(\mathbf{k}), J_2^0(\mathbf{k'}) \rangle = 8\pi \delta(\gamma - \gamma') F_2(\omega, T) \exp(-i(\alpha - \alpha')d) \frac{\sin((\alpha - \alpha')a_2)\sin((\beta - \beta')b_2)}{(\alpha - \alpha')(\beta - \beta')},$$

$$\langle J_1^0(\mathbf{k}), J_1^0(\mathbf{k'}) \rangle = 8\pi \delta(\gamma - \gamma') F_1(\omega, T) \frac{\sin((\alpha - \alpha')a_1)\sin((\beta - \beta')b_1)}{(\alpha - \alpha')(\beta - \beta')}.$$

Since $F_m(\omega, T) = 0$ where $\varepsilon_m = 1$ (in a vacuum) the correlations are zero. Also completed $\langle J_1^0(\mathbf{k}), J_2^0(\mathbf{k'}) \rangle = 0$. It is obvious that in the limit at $a_m, b_m \to \infty$ will be $\langle J_m^0(\mathbf{k}), J_m^0(\mathbf{k'}) \rangle = (2\pi)^3 \delta(\alpha - \alpha') \delta(\beta - \beta') \delta(\gamma - \gamma') F_m(\omega, T)$, i.e. in an infinite medium the spectral correlations are completely delta-correlated. However, this is not the case for finite bodies. We will be interested in spatial correlations:

$$\langle J_2(\mathbf{r}), J_2^0(\mathbf{r'}) \rangle = \delta(z - z') F_2^{02}(\omega, \beta, \beta', \gamma, T),$$

$$\langle J_2(\mathbf{r}), J_2(\mathbf{r'}) \rangle = \delta(z - z') F_2^2(\omega, \beta, \beta', \gamma, T).$$

They will be used to calculate correlations using the GF (12) representation. To do this, in similar expressions, using FG (13), perform two integrations over $\alpha$ and $\alpha'$ by the method of residue theory. As a result, we get

$$F_2^{02}(\omega, \beta, \beta', \gamma, T) = 4 F_2(\omega, T) \exp(-(\kappa + \kappa')d) \frac{\sinh((\kappa + \kappa')a_2)\sin((\beta - \beta')b_2)\xi_2 \Gamma(\gamma)}{(\kappa + \kappa')(\beta - \beta')[\kappa - (\xi_1 + \xi_2)\Gamma(\gamma)]},$$

$$F_2^2(\omega,\beta,\beta',\gamma,T) = 4|\xi_2\Gamma(\gamma)|^2 \cdot$$

$$\cdot \frac{\sin((\kappa+\kappa')a_1)\sin((\beta-\beta')b_1) + \exp(-(\kappa+\kappa')d)\sin((\kappa+\kappa')a_2)\sin((\beta-\beta')b_2)}{(\kappa+\kappa')(\beta-\beta')(\kappa-(\xi_1+\xi_2)\Gamma(\gamma))(\kappa'-(\xi_1+\xi_2)\Gamma(\gamma))^*}$$

Here we need to exclude a term that does not depend on the distance $d$:

$$F_2^2(\omega,\beta,\beta',\gamma,T) = 4|\xi_2\Gamma(\gamma)|^2 \frac{\exp(-(\kappa+\kappa')d)\sin((\kappa+\kappa')a_2)\sin((\beta-\beta')b_2)}{(\kappa+\kappa')(\beta-\beta')(\kappa-(\xi_1+\xi_2)\Gamma(\gamma))(\kappa'-(\xi_1+\xi_2)\Gamma(\gamma))^*} . \quad (26)$$

In these formulas $\Gamma^e_{zz}(\mathbf{k}) = \sqrt{\mu_0/\varepsilon_0}\,\Gamma(\gamma)G(\mathbf{k})$, $\Gamma(\gamma) = -i(k_0^2 - \gamma^2)/k_0$, $\xi_n = \sqrt{\mu_0/\varepsilon_0}\,\sigma_n$. Now we can write

$$\langle J_n(\mathbf{k}), J_m^0(\mathbf{k}')\rangle = 2\pi\delta(\gamma-\gamma')F_n^{0m}(\omega,\beta,\beta',\gamma,T),$$

$$\langle J_m(\mathbf{k}), J_n(\mathbf{k}')\rangle = \langle J_n(\mathbf{k}'), J_m(\mathbf{k})\rangle^* = (2\pi)\delta(\gamma-\gamma')F_m^n(\omega,\beta,\beta',\gamma,T),$$

There is a relation $F_n^{m^*} = F_m^n$. Thus, the induced current densities on different rods are not fully delta-correlated. Now you can write full correlations for the field components. Denoting the value (20) as $\langle E_x^0(a_1), E_x^0(a_1)\rangle = \Phi_{x2}(\mathbf{a},\mathbf{b},\omega,T)$, we get

$$\langle E_x(a_1), E_x(a_1)\rangle = \Phi_{x2}(\mathbf{a},\mathbf{b},\omega,T) + 2\mathrm{Re}(\Phi_{x2}^{02}(\mathbf{a},\mathbf{b},\omega,T)) + \Phi_{x2}^2(\mathbf{a},\mathbf{b},\omega,T).$$

Here $\mathbf{a} = (a_1, a_2)$, $\mathbf{b} = (b_1, b_2)$. The values $\Phi_{x2}^{02}$ and $\Phi_{x2}^{02}$ differ from $\Phi_{x2}$ by the fact that instead of the value $F_2$, the values $F_2^{02}(\gamma)$ and $F_2^2(\gamma)$ are included, respectively, but under the integral. Consider two other correlations of the electric field at $x = a_1$:

$$\langle E_y(a_1), E_y(a_1)\rangle = \Phi_{y2}(\mathbf{a},\mathbf{b},\omega,T) + 2\mathrm{Re}(\Phi_{y2}^{02}(\mathbf{a},\mathbf{b},\omega,T)) + \Phi_{y2}^2(\mathbf{a},\mathbf{b},\omega,T),$$

$$\langle E_z(a_1), E_z(a_1)\rangle = \Phi_{z2}(\mathbf{a},\mathbf{b},\omega,T) + 2\mathrm{Re}(\Phi_{z2}^{02}(\mathbf{a},\mathbf{b},\omega,T)) + \Phi_{z2}^2(\mathbf{a},\mathbf{b},\omega,T).$$

To calculate these values, we use integrals of the type

$$E_z^d(a_1, y, z) = \int_{V_2}\int_{-\infty}^{\infty}(k_0^2 - \gamma^2)\frac{\exp(-i\gamma(z-z') - i\beta(y-y') - \kappa|a_2 - x'|)}{8i\omega\varepsilon_0\pi^2\kappa} J_{2z}(\mathbf{r}')d\gamma d\beta d^3r',$$

and to calculate the correlations of the magnetic field, integrals of the form

$$H_x^d(a_1, y, z) = -i\int_{V_2}\int_{-\infty}^{\infty}\beta\frac{\exp(-i\gamma(z-z') - i\beta(y-y') - \kappa|a_2 - x'|)}{8\pi^2\kappa} J_{2z}(\mathbf{r}')d\gamma d\beta d^3r'.$$

For the definition $E_y^d$ we replace $(k_0^2 - \gamma^2)$ with $-\beta\gamma$ in the first integral, and for the definition $H_y^d$ we replace $-i\beta$ with $-\kappa$ in the second integral. We are interested in complete correlations of the type $\langle E_y, E_y\rangle = \langle E_y^0, E_y^0\rangle + 2\mathrm{Re}\langle E_y^0, E_y^d\rangle + \langle E_y^d, E_y^d\rangle$. In the first there is only $J_2^0$, in the second there is $J_2$, and in the cross-over – $J_2^0, J_2$ and $J_2, J_2^0$. Their integrands differ, respectively, by the multipliers $F_2(\omega,T)$, $F_2^2(\omega,T,\gamma)$, and $2\text{Re}(F_2^{02}(\omega,T,\gamma))$ by the multipliers constructed from the quantities $\gamma$, $\beta$, $\beta'$, $\kappa$, $\kappa'$, responsible for the corresponding derived components of the vector potential $A_z$, through which the components of the fields are defined. To determine the total running force, use the values of the type $\Phi_x^\pm(\mathbf{a},\mathbf{b},\omega,T) = \langle E_x^0(a_1), E_x^0(a_1) \rangle - \langle E_x^0(-a_1), E_x^0(-a_1) \rangle$, i.e. also consider the tension on the other side $x = -a_1$ of the rod. You should also write down the correlations $\Psi_x^\pm(\mathbf{a},\mathbf{b},\omega,T) = \langle H_x(a_1), H_x(a_1) \rangle - \langle H_x(-a_1), H_x(-a_1) \rangle$ and similarly $\Psi_y^\pm(\mathbf{a},\mathbf{b},\omega,T)$. Now the full linear force is there

$$f_x(d) = \frac{\mu_0}{2\pi} \int_0^\infty \left[ \Psi_y^\pm(\omega,T) - \Psi_x^\pm(\omega,T) + \frac{\varepsilon_0}{\mu_0} \left( \Phi_y^\pm(\omega,T) + \Phi_z^\pm(\omega,T) - \Phi_x^\pm(\omega,T) \right) \right] d\omega. \qquad (27)$$

Here, the factor 1/2 in the EMT and the factor 1/2 when averaging the quadratic values over the period are taken into account, and the spectral integral over positive frequencies is doubled. The correlations in (27) are taken on both sides $x = \pm a_1$, which is reduced simply to substitution $d_0^2 \to d_0^2(a_1) - d_0^2(-a_1) = d_0^\pm$. We have

$$d_0^\pm = \frac{4\exp(-(\kappa+\kappa'^*)d)\sinh((\kappa+\kappa'^*)a_1)\sinh((\kappa+\kappa'^*)a_2)}{(\kappa+\kappa'^*)}.$$

The force (27) does not depend on the $z$ coordinate (which fell out of the final result), but depends on $d$, and tends to zero at $d \to \infty$. However, the specified dependence on $d$ takes place only if $\kappa > 0$, and $\kappa' > 0$, i.e. $\gamma^2 + \beta'^2 > k_0^2$, and $\gamma^2 + \beta^2 > k_0^2$, which is what we originally meant by selecting the specified branches of the root and considering $\kappa$ and $\kappa'$ to be real. These are propagating along $z$ and evanescent along $x$ PP. Otherwise, when integrating by $x'$ instead of the decaying exponent $\exp(-(\kappa+\kappa')(x' \mp a_1))$ (always on the second body $x' > \mp a_1$) there is an oscillating exponent $\exp(-i(|\kappa|-|\kappa'|)(x'-a_2))$. The result of integration over $x'$ oscillates during integration over $\gamma, \beta, \beta'$ and does not contribute to the force. In addition, the integrals formally diverge. Therefore, the areas $\gamma^2 + \beta^2 < k_0^2$, $\gamma^2 + \beta'^2 < k_0^2$ should be excluded from the integration. These are the regions of fast radiated PP. For them, the rods are almost transparent, and the PP is almost not connected and does not interact. The transparency is due to the fact that the waves are radiated in all directions, and according to the condition of the problem $a_n \ll d$, $b_n \ll d$. Note that in (27) the is no singularity at $\beta' = \beta$. The convergence of the integrals provides an exponentially decaying factor $d_0^\pm$ for large $\gamma, \beta, \beta'$. Therefore, integrals can be calculated numerically.

In the case of a symmetric structure $a_n = a$, $b_n = b$, we have

$$d_0^\pm = \frac{4\sin^2((\kappa+\kappa')a)\exp(-(\kappa+\kappa')d)}{(\kappa+\kappa')}.$$

The given integrals have a complex dependence on $d$, as well as on $a$ and $b$. In order to establish such an analytical relationship, we consider the approximate calculation of integrals. Using substitution $\beta' = \beta + \tau$, we obtain functions for even and odd over $\beta'$ correlations:

$$f^\pm(\beta,\tau) = \frac{\sin^2(\tau b)}{\tau^2} \pm \frac{\sin^2((2\beta+\tau)b)}{(2\beta+\tau)^2} \approx \frac{1}{2\tau^2} \pm \frac{1}{2(2\beta+\tau)^2}.$$

The approximate result is valid for large $b$, when the oscillating cosines of the doubled argument can be ignored. However, this can only be used for extremely large sizes. In our case of small transverse dimensions, we should take

$$f^\pm(\beta,\beta') \approx b^2(1-b(\beta'-\beta)/3) \pm b^2(1-b(\beta'+\beta)/3).$$

In this expression, you can neglect the last term, or even the last two terms. In addition, the contribution of correlations involving odd function $f^-$ can be neglected compared to contributions with even function $f^+$. For small a, we have $d_0^\pm = 4a^2(\kappa+\kappa')\exp(-(\kappa+\kappa')d)$, $f^+ = 2b^2 - 2b^3\beta'/3$, i.e., the force in the roughest approximation is proportional to $(ab)^2$. For the approximate calculation of the force (27), we take into account only integrals with correlation $\langle J_2^0, J_2^0 \rangle$, in which we neglect the terms under which the integrals include the product $\beta\beta'$, i.e., the function $f^-(\beta,\beta')$, since it gives a significantly smaller contribution than $f^+(\beta,\beta')$. These are correlations $\langle E_y, E_y \rangle$ and $\langle H_x, H_x \rangle$. The contribution from $\langle E_x, E_x \rangle$ is significantly smaller than from $\langle E_z, E_z \rangle$, since the structure is not bounded on the z axis, and the size $a$ is small. Since we have neglected $\langle J_2, J_2^0 \rangle$, $\langle J_2, J_2 \rangle$, this means that the zero approximation of perturbation theory is used. Namely, the original equations for $J_n$ have the form of the Lippmann-Schwinger equations. We have obtained their exact spectral solutions, but they can be solved by the method of successive approximations. Zero approximation just means $J_2 = 0$. This approximation does not take into account the induced current, i.e. the influence of the diffraction field, and the first body is affected only by the fluctuation field of the second body. Then we have

$$f_x(d) \approx \frac{\mu_0}{2\pi}\int_0^\infty \left(\Psi_y^\pm(\omega,T) + \frac{\varepsilon_0}{\mu_0}\Phi_z^\pm(\omega,T)\right)d\omega. \qquad (28)$$

Let's combine the integrand in (28) into one integral and write it as

$$F_2(\omega,T)\int_{k_0}^\infty \left(k_0^2 - \gamma^2 + \frac{\gamma^4}{k_0^2}\right)d\gamma \int_{\beta_0}^\infty\int_{\beta_0}^\infty \frac{d_0^2(\gamma,\beta,\beta')f^+(\beta,\beta')d\beta d\beta'}{2\pi\kappa\kappa'}.$$

Using the small $a_1 = a_2 = a$ and $b_1 = b_2 = b$ approximation, we obtain

$$\frac{4(ab)^2 F_2(\omega,T)}{\pi} \int_0^\infty \left(k_0^2 - \gamma^2 + \frac{\gamma^4}{k_0^2}\right) d\gamma \int_{\beta_0}^\infty \exp(-\kappa d) \int_{\beta_0}^\infty \left(\frac{1}{\kappa} + \frac{1}{\kappa'}\right) \exp(-\kappa' d)\left(1 - \frac{2b\beta'}{3}\right) d\beta d\beta'.$$

We integrate by the parts over $\beta'$ with the replacement $\beta' = \sqrt{\tau^2 + \beta_0^2}$ and leave only the first terms with minimal negative powers of $d$. The integral is equal to $(d\beta_0)^{-1}[1+(d\kappa)^{-1}](1-2b\beta_0/3)\exp(-\beta_0 d)$. Similarly, integrating over $\beta$, we get the result

$$\frac{8(ab)^2 F_2(\omega,T)}{3\pi d^3} \int_{k_0}^\infty \left(k_0^2 - \gamma^2 + \frac{\gamma^4}{k_0^2}\right)\left(\frac{3 - 2b\sqrt{\gamma^2 - k_0^2}}{\gamma^2 - k_0^2}\right) \exp\left(-2d\sqrt{\gamma^2 - k_0^2}\right) d\gamma.$$

From this integral, we must exclude the divergent term corresponding to the summand $\gamma^4/k_0^2$. It was obtained due to the approximations used at the expense of discarded integrals, since the original integral converges. Making a replacement $u^2 = \gamma^2 - k_0^2$, we get

$$\frac{8(ab)^2 F_2(\omega,T)}{3\pi d^3 k_0^2} \int_0^\infty \sqrt{u^2 + k_0^2}\, (3 - 2bu)\exp(-2ud) u\, du.$$

We calculate this integral in parts:

$$\int_0^\infty \sqrt{u^2 + k_0^2}\,(3u - 2bu^2)\exp(-2ud) du = \frac{3k_0}{(2d)^2} + \int_0^\infty \frac{\exp(-2ud)}{(2d)^2}\left(\sqrt{u^2 + k_0^2}(3u - 2bu^2)\right)'' du.$$

Integrating again and omitting the remaining integral, we get the final result

$$f_x(d) \approx \frac{(ab)^2}{\pi^3 d^5 c}\left(1 - \frac{2b}{3d}\right) \int_0^\infty \varepsilon_2''(\omega)\Theta(\omega,T) d\omega. \tag{29}$$

The asymptotic result (29) has dimension N/m, i.e. it is a linear force. Strict numerical integration (27) gives a complex dependence on the dimensions and on the distance $d$. In principle, the expression (29) can be modified by another integration in parts leading to a convergent integral. In this case, the parenthesis should be replaced with $(1 - 2b/(3d) + b/(3k_0 d^2))$ and inserted under the integral. However, such a modification can be valid for very large $d$, for which, with the spectrum used, it is fulfilled $k_0 d > 1$. Formulas (27) and (29) are valid for $d \gg a,b$. The result (29) significantly depends on the frequency dependence of the DP. It is convenient to take $\varepsilon_n = \varepsilon$ and use the Lorentz formula

$$\varepsilon(\omega) = 1 + \sum_{n=1}^N \frac{\omega_n^2}{\omega_{0n}^2 - \omega(\omega - i\omega_{cn})}. \tag{30}$$

In (30), you need to know the forces of the oscillators $\omega_n^2$, as well as the influence of the internal field on them. In the simplest case, $N=1$, we have $\varepsilon''(\omega) = \omega_1^2 \omega \omega_{c1}/[(\omega_{01}^2 - \omega^2)^2 + \omega^2 \omega_{c1}^2]$. However, the integral (29) can be analytically calculated only for $T = 0$, when $\Theta(\omega,0) = \omega\hbar/2$. We do not give this result. For metal $\omega_{01}=0$, and we get the Drude formula. Adding to it several terms in (30), we obtain the Drude-Lorentz formula, which takes into account the transitions in the lattice atoms. In this case, instead of zero at zero frequency, we have a pole, and divergence (29) at low frequencies. It is associated with an infinite length of rods. To exclude it, you can put the frequency very small $\omega_{01}=c\pi/L$, but not equal to zero. It corresponds to a wavelength of the order of the maximum size of the structure *L*. This is the Drude-Smith formula [19,20]. The low frequency means that the conduction electron is not completely delocalized from the atom, which is the case for metal rods (clusters) of not too long length. At the same time, for wavelengths that significantly exceed the size of the metal body, i.e. $\lambda \to \infty$, when there is a tendency $\varepsilon''(\omega)$ to zero.

The divergence at zero for an ideal boundless metal structure can be explained as follows. The divergence is related to the representation of the electric field in terms of the vector potential $\mathbf{E}=-i\omega\mu_0\mathbf{A}+(i\omega\varepsilon_0)^{-1}\nabla(\nabla\cdot\mathbf{A})$, where $\nabla\cdot\mathbf{A}=-i\omega\varepsilon_0\varphi^e$ and with the assumption of an infinite structure along the *z* axis. When striving $\omega \to 0$, we have $\mathbf{E}\approx(i\omega\varepsilon_0)^{-1}\nabla(\nabla\cdot\mathbf{A})=-\nabla\varphi^e$. To describe the electric field only with a scalar potential $\varphi^e$, the divergence of the vector potential must tend to zero in proportion to the frequency. In our case $\mathbf{J}=\mathbf{z}_0 J_z$, $\mathbf{A}=\mathbf{z}_0 A_z$, there is constant over the cross section of the charge density $\rho_V(z)=i\omega^{-1}\partial_z J_z(z)$. It creates the electric potential and the field. However, in statics $\partial_z J_z(z)=0$, and charges can only occur at infinity. From the general integral equations, one can obtain a representation of the electric field

$$\mathbf{E}(\mathbf{r})=-\varepsilon_0^{-1}\nabla\int_V G(0,\mathbf{r}-\mathbf{r}')\rho_V(\mathbf{r}')d^3r'.$$

When moving to the spectral representations for the components, we represent its component in the form

$$E_x(\mathbf{k})=\frac{i\alpha\rho_V(\mathbf{k})}{(2\pi)^3\varepsilon_0\mathbf{k}^2}=\frac{i\gamma\alpha J_z(\mathbf{k})}{(2\pi)^3\omega\varepsilon_0\mathbf{k}^2},$$

and similarly for other components. It follows that at low frequencies, the values $\gamma$ and $\omega$ cannot be independent and must consistently tend to zero. In field correlations, this can be done by replacing $\gamma^2 \to \gamma^2\omega^2/(\omega^2+\omega_0^2)$, where $\omega_0^2$ is a certain characteristic frequency. In another way, this means that the correlations $\langle J_z(\omega,\mathbf{k}),J_z(\omega,\mathbf{k})\rangle$ should tend to zero at $\omega \to 0$. Indeed, any real structure is finite, i.e. there are zero boundary conditions for the current. In such a structure, near the boundaries and correlations of fluctuating currents should tend to zero. In a finite structure with size *L=2l*, a polariton standing wave $J_z(z)\sim\cos(k_0 z)$ is excited at low frequencies. From the condition $J_z(\pm l)=0$, we find the specified frequency $\omega_0=c\pi/L$. For the final structure, you can define the total strength. In this case, you can use the expansion $J_z(\mathbf{r})$ and $J_z^0(\mathbf{r})$ in the Fourier series and the solution of the corresponding integral equations. However, this approach requires determining in the general case an infinite number of expansion coefficients from the solution of an infinite system of linear equations, and is significantly more complex. The spectra $J_z(\omega,\mathbf{k})$ and $J_z^0(\omega,\mathbf{k})$ are determined on the final carriers and contain the specified coefficients. According to the Wiener-Paley theorem on integer functions [21], these spectra decrease at infinity (at $k^2 \to \infty$). However, at $\omega \to 0$, $\gamma \to 0$, there must be $\langle J_{nz}^0(\omega,\mathbf{k}), J_{nz}^0(\omega,\mathbf{k})\rangle \to 0$, and not $\langle J_{nz}^0(\omega,\mathbf{k}), J_{nz}^0(\omega,\mathbf{k})\rangle \to F_n(0,T) = \sigma_n'(0)k_B T/\pi$, as was assumed for an infinite structure. Otherwise, the correlations will not vanish outside the structure. Note that we have actually imposed such an inversion in transverse coordinates by calculating spatial correlations over finite transverse regions. This difficulty can be circumvented by replacing $F_n(\omega,T) \to F_n(\omega,T)\omega^2/(\omega^2 + (c\gamma_0)^2)$, where $\gamma_0 = \pi/L$. In fact, this means a more rigorous consideration of the variance of correlations, taking into account the spatial configuration. Note also that in the case of a dielectric rod, divergence does not occur, since the imaginary part of the permittivity has zero at zero frequency, in contrast to the pole for metals and plasma. The applicability of the Drude formula at low frequencies has been questioned in a number of papers due to a possible violation of the Nernst thermal theorem (the third principle of thermodynamics) [22, 23]. It should also be noted that the infinite limit for the frequency integral should also be bounded. First, this is the region of the absorption spectrum of the substance. Secondly, these are wavelengths of the order of the discreteness of the structure and less, when the substance is no longer described as a continuous medium. Third, it is the frequency range when the properties of the quantum vacuum change significantly, i.e. when energy of quanta $\hbar\omega$ it is comparable to the polarization energy of the vacuum $2m_e c^2$. The first two conditions occur much earlier than the third.

When using DP (30) in the case of sufficiently narrow spectral lines, the force (29) can be calculated approximately, since the value $\omega_{cn}$ gives the half-width of the spectral line:

$$f_x(d) \approx \frac{2(ab)^2}{\pi^3 d^5 c}\left(1 - \frac{2b}{3d}\right)\sum_{n=1}^{N} \varepsilon''(\omega_{0n})\Theta(\omega_{0n},T)\omega_{cn}. \tag{31}$$

Here you should take:

$$\varepsilon''(\omega) = \sum_{m=1}^{N} \frac{\omega_m^2 \omega \omega_{cm}}{(\omega_{0m}^2 - \omega^2)^2 + \omega^2 \omega_{cm}^2}. \tag{32}$$

Substituting (32) into (31) results in a double sum. However, for very narrow and sufficiently spaced spectral lines, the contributions at $m \neq n$ can be neglected by taking

$$f_x(d) \approx \frac{2(ab)^2}{\pi^3 d^5 c}\left(1 - \frac{2b}{3d}\right)\sum_{n=1}^{N} \frac{\omega_n^2}{\omega_{0n}}\Theta(\omega_{0n},T). \tag{31}$$

If the entire spectrum lies in the region where $\hbar\omega \gg 2k_B T$ (which for good dielectrics is usually the case for room temperature), then $\Theta(\omega_{0n}, T) = \hbar\omega_{0n}/2$, and we get an interesting result:

$$f_x(d) \approx \frac{\hbar(ab)^2}{\pi^3 d^5 c}\left(1 - \frac{2b}{3d}\right)\sum_{n=1}^{N}\omega_n^2.$$

It shows that the force does not depend on the resonant frequencies and is determined by the strength of the oscillators and their concentration. The same result holds for $T = 0$. It shows that the force does not depend on the resonant frequencies and is determined by the strength of the oscillators and their concentration. The same result holds for $T = 0$. For high temperatures, when the energy $k_b T$ significantly exceeds the energy spectrum of the substance,

$$f_x(d) \approx \frac{2(ab)^2 k_b T}{\pi^3 d^5 c}\left(1 - \frac{2b}{3d}\right)\sum_{n=1}^{N}\frac{\omega_n^2}{\omega_{0n}}.$$

This case can be realized when the substance of the cylinders is in the plasma state.

The obtained results can be used to estimate the dispersion forces in different structures, for example, for two CNTs. CNTs are characterized by a surface conductivity tensor $\hat{\varsigma}_n = (\varsigma_{n\varphi\varphi}, \varsigma_{nzz})$. We consider the CNTs parallel, and their radii small compared to the distance: $r_n \ll d$. In this case, you can ignore the azimuthal current and use only the components $\varsigma_{nzz}$. For metallic and semiconductor CNTs, the dynamic (frequency) conductivities depending on the radius, chirality, and frequency are given in a number of papers, for example, in [24]. A rigorous solution of the problem of dispersion forces requires its consideration in cylindrical coordinates using cylindrical functions and with a transition to local coordinate systems, which leads to complex results. Replacing the surface current density with a volume one (i.e., uniformly distributed over the entire CNT cross-section), we obtain the specific conductivities from the condition $2\pi r_n \varsigma_{nzz} = \pi r_n^2 \sigma_n$, or $\sigma_n = 2\varsigma_{nzz}/r_n$. Let's put $a_n = b_n$ and replace each CNT with an effective square cylinder with the size $a_n = r_n\sqrt{\pi}$ filled with a dielectric with DP $\varepsilon_n = 1 - i2\xi_{nzz}/(\omega\varepsilon_0 r_n)$. In this model, the transverse polarization currents are not automatically taken into account. Next, we use the above approach.

Finally, consider the question of PP. They arise when the structure is excited and propagate along it with attenuation. Formally, the dispersion equations for PP are obtained as solutions of homogeneous equations (at $\mathbf{J}^0 = 0$). In our configuration, this leads to complex $\gamma$. The excitation of an open structure requires taking into account the entire continuous and discrete spectra and cannot lead to the appearance of a single mode [25]. Radiation arise both quick radiative waves and the surface PP with different frequencies and amplitudes. Accordingly, the excitation is a non-stationary process.

However, it can be quasi-stationary with the excitation of a predominantly single mode. After the excitation ceases, the radiation modes are highlighted, and the surface PPs propagate with attenuation and, accordingly, with a decrease in amplitudes [25]. After a while, the PP remains with mostly low attenuation. Consider this particular mode. For it, the equations are homogeneous, i.e. it should be put $J_n^{0'} = 0$. There are infinitely many such modes, depending on how the field depends on the transverse coordinates. You should use Fourier expansions, solve the corresponding equations, and derive the dispersion equations from them. In general, depending on the frequency, a particular PP can be slow surface or fast radiated. The spectral problem can be formulated as finding the propagation constant $\gamma$ for given transverse wavenumber $\kappa_{mn} = \sqrt{\alpha_m^2 + \beta_n^2}$ and $k_0$. Solutions in the form of coupled waves (polaritons) in dielectric waveguides or coupled PP in the case of metal rods are interesting. The problem should be solved numerically, but it is possible to obtain an approximate associated dispersion equation, which includes multipliers of the type $\exp\left(-2d\sqrt{\gamma^2 + \beta_n^2 - k_0^2}\right)$. Such a dispersion equation decays at a large distance into two uncoupled PP. The analytical form of the dispersion equation can be obtained for plates [26].

In this paper, we propose a model for determining the Casimir-Lifshitz force for two different long parallel dielectric rods, as well as for two CNTs. The model uses the fluctuation-dissipation theorem, is based on the representation of fields through the GFs using the permittivity and on the determination of the force using the EMT in a vacuum. The result is presented as a spectral integral over the frequency spectrum and over the spatial spectra. The spatial spectra should be limited from below to the values at which the PP becomes rapidly radiated, and the frequency to a certain maximum value at which the DP is close to one, and the imaginary part of the DP is zero. Using the Lorentz formula leads to a convergent frequency integral. For metallic structures, it is proposed to use the Drude-Smith formula for convergence. An approximate expression for the linear force for small transverse dimensions is obtained, which is presented as an analytical formula with the main contribution of the order of the inverse fifth power of the distance. For the Lorentz dispersion, the spectral integral is calculated approximately analytically. It is shown that for $2k_B \ll \hbar\omega_{0n}$ the result does not depend on the resonant frequencies $\omega_{0n}$.

Taking into account the induced current densities leads to additional integrals, which can also be estimated or calculated, and by virtue of (25), the force dependence includes $\varepsilon_1''$, $\varepsilon_2''$, $\varepsilon_1$, and $\varepsilon_2$. The used approximation loses its meaning at distances of the order of or less than the transverse dimensions. Restrictions on small distances are generally inherent in the formulas of dispersion forces: at low wavelengths $\lambda \sim d$ the structure becomes transparent, and the forces cease to act, which was noted still by Casimir. However, usually the discreteness of the structure takes effect even earlier, and the van der Waals repulsive forces arise due to the overlap of the electron shells. The rigorous consideration of bodies of finite and complex configuration requires the numerical solution of the electrodynamic equations, which are reduced to the inversion of large-order matrices, i.e., to the expression of approximations of fields and currents through fluctuations and to the calculation of their correlations [16].

The work was supported by the Ministry of Education and Science of the Russian Federation as part of the state task (project No. FSRR-2020-0004).